\documentclass{ws-mpla}

\usepackage{amsmath}

\def\lsim{\raise0.3ex\hbox{$<$\kern-0.75em\raise-1.1ex\hbox{$\sim$}}}
\def\gsim{\raise0.3ex\hbox{$>$\kern-0.75em\raise-1.1ex\hbox{$\sim$}}}

\begin{document}

\markboth{Carlos A. Salgado}
{Medium Induced gluon radiation and jet quenching in heavy Ion collisions.}

%
\catchline{}{}{}{}{}
%

\title{MEDIUM-INDUCED GLUON RADIATION AND\\
JET QUENCHING IN HEAVY ION COLLISIONS.
}

\author{\footnotesize CARLOS A. SALGADO}

\address{Theory Division, CERN\\ CH-1211 Geneva 23, Switzerland\\
carlos.salgado@cern.ch}

\maketitle


\begin{abstract}
In this brief review, I summarize the new developments 
on the description of gluon radiation by energetic quarks traversing a medium
as well as the observable consequences in high-energy heavy ion collisions. 
Information about the initial state is essential for a 
reliable interpretation of the experimental results and will also be reviewed.
 Comparison with experimental data from RHIC and
expectation for the future LHC will be given.


\end{abstract}

\ccode{PACS Nos.: 25.75.-q, 12.38.Mh, 24.85.+p}

\section{Introduction}

The experimental program on high energy heavy ion collisions attempts to study
the behavior of QCD matter under extreme conditions. The original, and
still most important goal, is the creation and characterization 
of the quark-gluon plasma,
a thermalized state of deconfined quarks and gluons that could be
the form of matter 
of the whole Universe only several $\mu s$ after the Big-Bang. 
The study of high-$p_t$ processes as probes of the produced medium starts with 
the seminal work of J.D. Bjorken in 1982 \cite{bjorken}. 
The idea was that if a medium is
produced in a collision, the high-$p_t$ particles produced inside the medium
in the initial stage would loss energy (and eventually thermalize)
when escaping it.
The arguments in \cite{bjorken} were based on
elastic scattering, and the loss turned out to be too small.
Later refinements \cite{wang,bdmps,z}
propose  the medium-induced gluon radiation 
as the dominant source of energy loss. 

Twenty years
latter, in the heavy ion collider era started with RHIC, these
effects could be measured for the first time.
The new experimental facts coming from RHIC  when comparing
central AuAu with pp collisions are the following: the suppression of particles
with high-$p_t$ 
\cite{Adler:2003qi,Adams:2003kv,Arsene:2003yk} 
(independent of the particle species
for $p_t\gsim 4$ GeV) and the total extinction of the signal associated to a
high-$p_t$ particle in the backward hemisphere \cite{Adler:2002tq}
(back-to-back correlations).
Together with this, experimental data in dAu collisions find an enhancement
of high-$p_t$ particle production 
\cite{Arsene:2003yk,Adler:2003ii,Adams:2003im,Back:2003ns}
(the so-called Cronin effect) and back-to-back
correlations of the same magnitude as the ones measured in pp collisions
\cite{Adams:2003im}.
These effects point to a strong interaction of the high-$p_t$ particle with 
the (dense) produced medium in agreement with the {\it jet quenching} scenario.
In the following we will present the general formalism, based on collinear
factorization, in which most of the present
calculations are based as well as  some comparison with 
experimental data.

To leading order in perturbative QCD, high-$p_t$ hadroproduction in
proton--proton collisions is described by the factorization formula
 (see e.g. \cite{Eskola:2002kv}):
\begin{eqnarray}
E\frac{d\sigma^h}{d^3p}&=& K(\sqrt{s})\int dz dx_1 dx_2
\frac{\hat s}{\pi z^2}\delta(\hat s+\hat t+\hat u)\times\nonumber \\
 &&\times\sum_{i,j}\, f^A_i(x_1,Q^2)\, f^B_j(x_2,Q^2)\,
\frac{d\sigma^{ij\to k}}{d\hat t} \,
D_{k\to h}(z,Q^2),
\label{eqpqcd}
\end{eqnarray}

\noindent
where $f^A_i$, $f^B_j$ are the proton parton distribution functions (PDF),
$d\sigma^{ij\to k} / d \hat t$ is the partonic cross section and
$D_{k\to h}(z,Q^2)$ describes the fragmentation of a parton $k$ into the
hadron $h$ carrying a fraction $z$ of the momentum. The description of the 
experimental data
by (\ref{eqpqcd}) is very reasonable
\cite{Eskola:2002kv,Vitev:2002aa,Levai:2003at}.

In the case of heavy ion collisions, both the initial and, possibly, the final 
state are different. Indeed, the nuclear parton distributions are different
from those in free protons and the eventually produced medium would modify
the fragmentation. The knowledge of the PDF for bounded nucleons gives the
baseline for the final state effects which would provide 
the information about the
medium. So, the first goal is to obtain these nuclear PDF from experimental data
in different processes.

The review is organized as follows: in next section we give a description of 
the initial state effects in terms of nuclear modifications of PDF; Section
3 describes the medium-induced gluon radiation spectrum, which is the main
part of the present work; in Section 4 some applications are discussed, both for
inclusive particle production (where a comparison with RHIC data is possible)
as well as for the more differential case of jet observables. 
In the last two sections we comment on different approaches and give our 
conclusions.

\section{Initial state effects: shadowing}

Nuclear and free proton PDF -- $f_i^A$ and $f_i^p$ respectively -- are normally
related by the ratio $R_i^A$

\begin{equation}
f_i^A(x,Q^2)=R_{i}^A(x,Q^2)f_i^N(x,Q^2)\, .
\label{eqSalgado1}
\end{equation}
For the corresponding ratio of the structure function $F_2$ 
%
several different regions have been measured, as shadowing ($R_{F_2}^A <1$) at
small values of $x$, antishadowing ($R_{F_2}^A >1$)
for intermediate $x$ and EMC ($R_{F_2}^A <1$)
and Fermi motion for large $x$. In this way, a similar structure is expected
for the nuclear PDF $f_i^A(x,Q^2)$

The proton PDF are usually obtained in global fits to experimental data in 
well established DGLAP analysis\cite{Martin:2001es,Pumplin:2002vw}.
The main difficulty in applying the
same method to the nuclear case is the lack of experimental data. In this
section we present two sets of nuclear PDF 
(EKRS \cite{Eskola:1998df} and HKM \cite{Hirai:2001np})\footnote{By the
time this review was finished a new analysis was published 
\cite{deFlorian:2003qf}. 
Quantitative differences appear for the gluons when compared with
EKRS but the qualitative features are in agreement with this set. 
(See Ref. \cite{deFlorian:2003qf} for further details and the effect of these
corrections on the high-$p_t$ $\pi^0$ yields measured at RHIC.
See also \cite{Frankfurt:2002kd} for a related approach).}
and comment about 
the experimental constraints to the different flavors, specially those for
gluons. 

\subsection{EKRS analysis of nuclear PDF}

The goal of nuclear DGLAP analyzes is to obtain a set of nuclear
PDF following the procedure of the proton case. Namely, fixing the initial
parton distributions at a $Q_0^2\gg \Lambda_{QCD}^2$ and evolving them to
larger $Q^2$ values by DGLAP equations. The comparison with data would
fix the free parameters in the initial condition.
In practice, what is usually
done is to obtain the initial {\it ratios}, $R_i(x,Q^2_0)$
for different partons $i$, and use a known set of proton PDF (as MRST
\cite{Martin:2001es}, CTEQ \cite{Pumplin:2002vw},
etc...) to obtain the nuclear PDF. 
In the EKRS analysis, data on nuclear $F_2$, and DY measured in pA collisions
is used.
Further constraints are momentum and baryon number sum rules.
At the initial scale, the ratios for valence $R_V(x,Q^2_0)$
(same for $u_V$ and $d_V$), sea $R_S(x,Q^2_0)$
(same for $\bar u$, $\bar d$ and $\bar s$)
and gluons $R_g(x,Q^2_0)$
are obtained
in the following way:

\noindent
$\bullet$ At large values of $x$ ($x\gsim 0.3$), $R_{F_2}$ data is used to
fix the valence quarks ratio $R_V$. 
Both $R_S$ and $R_g$ are not constrained, so  
they are assumed to be equal to $R_V$. 

\noindent
$\bullet$ At intermediate values of $x$ ($0.04\lsim x\lsim 0.3$) both DIS and
DY data constrain the ratios $R_V$ and $R_S$. 
Baryon number sum rule imposes also
constraints to valence ratio. In this region, the gluon ratio is fixed by
momentum sum rule,  
with the help of NMC data \cite{Arneodo:1996ru} to fix the value of $x$ where 
$R_g(x)$=1 (see below). This produces a large gluon antishadowing.

\noindent
$\bullet$ At small values of $x$ ($x\lsim 0.04$),
$R_S=R_{F_2}$ and a saturation of $R_{F_2}$ is assumed for $x\lsim 10^{-3}$;
$R_g=R_{F_2}$ is taken
for the very small $x$ values -- notice that evolution modifies this equality;
$R_V$ is fixed by baryon number sum rule.

\begin{figure}[tb]
\begin{center}
\includegraphics[width=13cm]{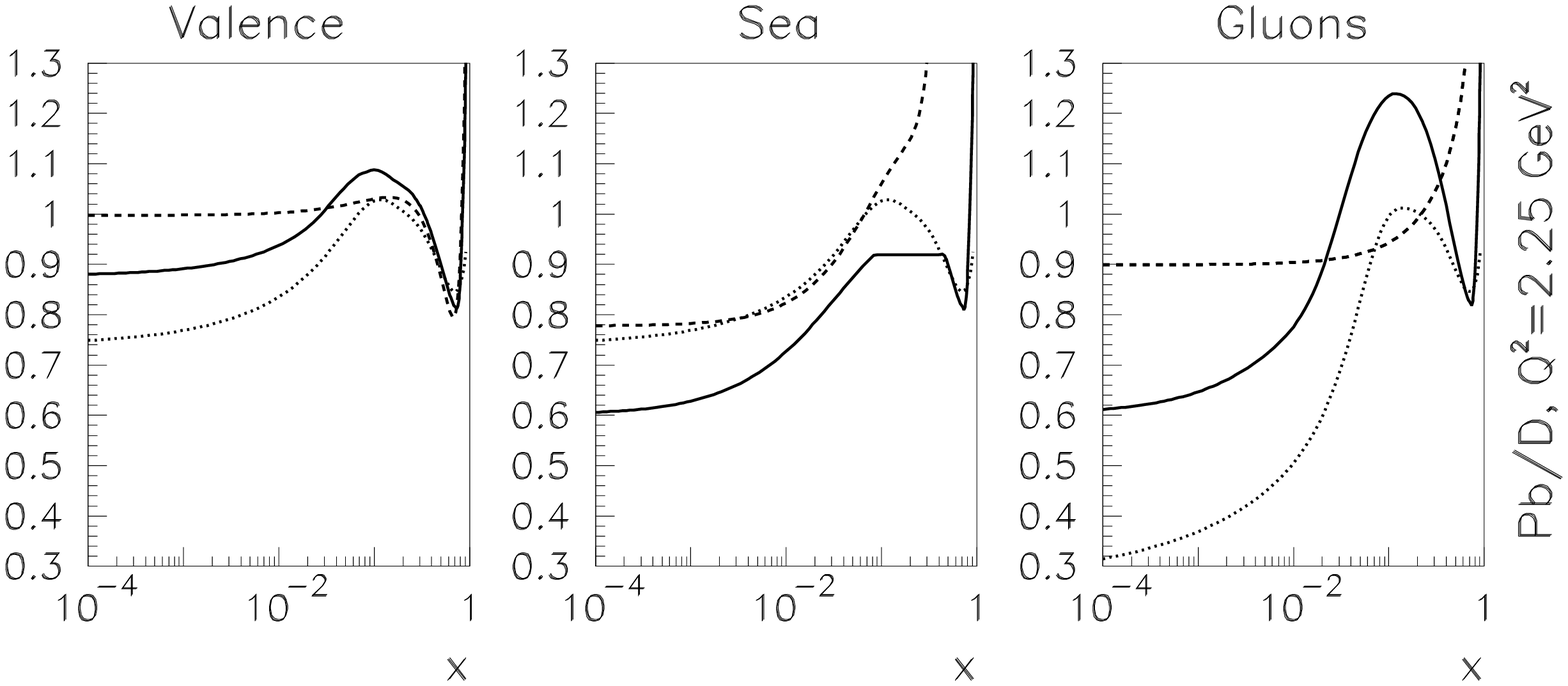}
\caption[a]{EKS98 (solid lines) and HKM (dashed lines) initial conditions for
the ratios of valence, sea and gluon PDF of Pb over deuterium. Also shown are
the corresponding ratios for the new HIJING parametrization (dotted lines).
}
\end{center}
\label{figSalgado2}
\end{figure}

Once the initial conditions are known, LO-DGLAP  evolution is
performed, and the parameters of the initial conditions fixed by comparing 
to data at different
values of $Q^2$. The initial conditions obtained by this method are plotted
for a Pb nucleus
in Fig. \ref{figSalgado2} and compared with 
HKM \cite{Hirai:2001np}.
The main differences come from the fact that HKM do not use
data on Drell-Yan ~\cite{Alde:im} with nuclei (essential to constrain
valence and sea quarks at intermediate $x$) nor the 
$Q^2$-dependent data measured by NMC \cite{Arneodo:1996ru}.

\subsection{Constraints for gluons}

\begin{figure}[tb]
\begin{center}
\includegraphics[width=10cm]{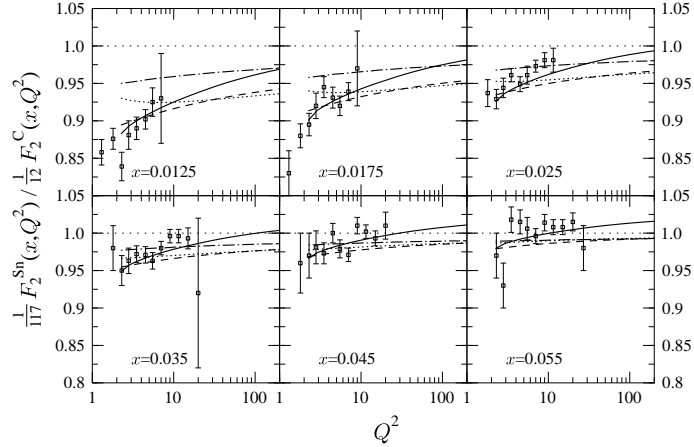}
\vskip -0.5cm
\caption[a]{{$Q^2$-evolution of the ratios
of $F_2$ in Sn and C for different initial conditions
EKS98 \cite{Eskola:1998df} (solid lines), HKM \cite{Hirai:2001np}
(dotted-dashed lines), HPC \cite{HPC} (dashed lines) and
new HIJING \cite{Li:2001xa} (dotted lines)
compared with the NMC results.
}}
\label{figSalgado3}
\end{center}
\end{figure}

At LO, the gluon distribution does not directly contribute to the DIS or DY
cross sections. 
It instead drives the
$Q^2-$evolution of all other flavors:
at small
values of $x$, DGLAP at LO gives $\partial F_2^{p(n)}/\partial \log Q^2
\sim  xg(2x,Q^2)$. So, for the ratios,
\begin{equation}
\partial R_{F_2}^A(x,Q^2)/\partial \log Q^2
\propto
 \left\{R_g^A(2x,Q^2)-R_{F_2}^A(x,Q^2)\right\} .
\label{eqSalgado4}
\end{equation}
In order to obtain a positive $\log Q^2$-slope, as measured by the NMC
Collaboration \cite{Arneodo:1996ru} (see Fig. \ref{figSalgado3})
a very strong gluon shadowing for $x\gsim 0.01$
is not allowed by (\ref{eqSalgado4}).
To quantify this statement \cite{Eskola:2002us}, 
we have applied DGLAP evolution
to different initial conditions and compare with the NMC data, the comparison
is done in  Fig. 
\ref{figSalgado3}.
The slopes
reflect the gluon distribution, in particular, the negative slopes
obtained when taken new HIJING parametrization \cite{Li:2001xa} indicate that
the very strong shadowing for gluons is
in disagreement with data. (This is, also, in agreement with the new analysis
\cite{deFlorian:2003qf}).

Summarizing, in order to use the collinear factorization formula 
(\ref{eqpqcd}) 
a set of nuclear PDF is needed. These nuclear PDF can be constrained
by DIS and DY experimental data and evolved by DGLAP equations. In this 
framework, a strong gluon shadowing for $x\gsim 0.01$ is not supported
by present data. EKRS
parametrization 
gives\cite{Eskola:2002kv}, 
for RHIC at $y\sim 0$, a moderate enhancement in the intermediate region
of $p_t$,  but interestingly, in agreement 
with the increase in the $\pi^0$ yield measured in dAu collisions
\cite{Adler:2003ii}. 

\section{Final-state effects: medium-induced gluon radiation}

The 
medium-induced distribution of gluons of energy $\omega$
radiated off an initial hard parton has been computed by several methods and
approximations~\cite{bdmps,z,glv,Wiedemann:2000za}. 
They can be obtained as particular cases of  the general 
$k_t$-differential spectrum~\cite{Wiedemann:2000za,Salgado:2003gb}
\begin{eqnarray}
  \omega\frac{dI}{d\omega\, d{\bf k}}
  &=& {\alpha_s\,  C_R\over (2\pi)^2\, \omega^2}\,
    2{\rm Re} \int_{0}^{\infty}\hspace{-0.3cm} dy_l
  \int_{y_l}^{\infty} \hspace{-0.3cm} d\bar{y}_l\,
   \int d^2{\bf u}\,
  e^{-i{\bf k}_t\cdot{\bf u}}   \,
  e^{ -\frac{1}{2} \int_{\bar{y}_l}^{\infty} d\xi\, n(\xi)\, \sigma({\bf u})}\,
  \times 
\nonumber\\
  &\times&{\partial \over \partial {\bf y}}\cdot
  {\partial \over \partial {\bf u}}\,
  \int_{{\bf y}={\bf r}(y_l)}^{{\bf u}={\bf r}(\bar{y}_l)}
  \hspace{-0.5cm} {\cal D}{\bf r}
   \exp\left[ i \int_{y_l}^{\bar{y}_l} \hspace{-0.2cm} d\xi
        \frac{\omega}{2} \left(\dot{\bf r}^2
          - \frac{n(\xi)\, \sigma({\bf r})}{i\,2\, \omega} \right)
                      \right]\, .
    \label{eqspec}
\end{eqnarray}
Here, $C_R$=$C_F$=4/3 for quarks and $C_R$=$C_A$=3 for gluons.
Medium properties enter (\ref{eqspec}) via the product of the
medium density $n(\xi)$ of scattering centers times the dipole
cross section $\sigma({\bf r})$ which measures the
interaction strength of a single elastic scattering.
The solution for a general $n(\xi)\sigma({\bf r})$ is unknown, and 
two approximations have been studied up to now: the multiple soft 
scattering limit, $n(\xi)\, \sigma({\bf r})
\approx \frac{1}{2} \hat{q}(\xi)\, {\bf r}^2$, in which the path integrals
reduce to those of a harmonic oscillator and can be solved analytically
-- this is the approximation used by the BDMPS group \cite{bdmps} and Zakharov
\cite{z};
the single hard scattering limit, which consists in a series expansion in
$n(\xi)\sigma({\bf r})$, 
where $\sigma({\bf r})$ is modeled by a Yukawa potential with Debye
screening mass $\mu$ -- this is the approximation used by the GLV group
\cite{glv}.

\begin{figure}[tb]
\begin{center}
\includegraphics[width=8cm]{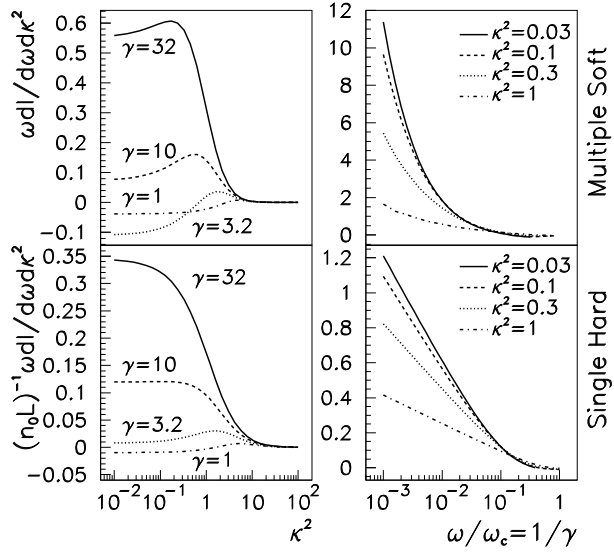}
\caption[a]{{
The gluon energy distribution
(\protect\ref{eqspec}) as a function of the rescaled gluon energy
$\omega/\omega_c$ and the rescaled gluon transverse momentum
$\kappa$, see eq. (\ref{eqvar}). ($\alpha_s$=1/3 has been taken).
}}
\end{center}
\label{figdidwdkt}
\end{figure}

Eq.~(\ref{eqspec}) implies a
one-to-one correspondence between the average energy loss
of the parent parton, and the transverse momentum broadening
of the associated gluon radiation, as argued in
Ref.~\cite{Baier:1996sk}. In Fig. \ref{figdidwdkt} 
we present the numerical results of 
$\omega dI/d\omega\, d{\bf k}$ for quarks 
in the two approximations. 

Most of the qualitative properties of the 
medium-induced gluon radiation spectrum can be understood by 
coherence arguments. Let us concentrate on the multiple
soft scattering approximation, the same arguments hold for the single hard
with the change $\hat q \ \rightarrow\ \mu^2/L$.
For a gluon emitted with energy $\omega$ and transverse momentum
$k_t$, the phase and the gluon formation time are
\begin{equation}
  \varphi = \Bigg\langle \frac{k_t^2}{2\omega}\, \Delta z \Bigg\rangle
\Longrightarrow
l_{coh}\sim\frac{\omega}{k_t^2}\, .
\end{equation}
\noindent
The medium is characterized by the transport coefficient $\hat q
\simeq\frac{\mu^2}{\lambda}$, giving the average transverse momentum $\mu^2$ 
transfered from the medium to the gluon per
mean free path $\lambda$. So, $k_t^2\sim \mu^2 l_{\rm coh}/\lambda$, when 
$l_{\rm coh}$ reaches the length of the medium $L$ one has $k_t^2\sim
\hat q L$. So, this is a maximum for the $k_t$ of the emitted gluon. If one
defines
\begin{equation}
\kappa^2=\frac{k_t^2}{\hat qL} \ , \ \omega_c=\frac{1}{2} \hat qL^2
\hspace{0.5cm} \Bigg[
\kappa^2=\frac{k_t^2}{\mu^2} \ , \ \omega_c=\frac{1}{2} \mu^2L
\ \ {\rm For\ Single\ Hard}\Bigg]\ ,
\label{eqvar}
\end{equation}
\noindent
the phase for $\Delta z=L$ is $\varphi\sim \kappa^2\  \omega_c/\omega$. The
radiation can only be formed when $\varphi\gsim 1$, so
a suppression of the radiation appears when $\kappa^2\lsim\omega_c/\omega$. The 
plateau at small values of $\kappa$ for fixed $\omega/\omega_c$ in
Figure \ref{figdidwdkt}
is due to these coherence effects. Moreover, at large values of
$\omega\gsim\omega_c$ the spectrum is also suppressed. This is the well known
LPM suppression first discussed by the BDMPS group \cite{bdmps}. These features
are characteristic of QCD as the multiple scattering is performed by the
(eventually) emitted gluon in the high-energy approximation. In the case
of QED, the photon does not interact and the relevant phase contains now
the energy of the electron instead of $\omega$.

\begin{figure}[tb]
\begin{center}
\includegraphics[width=10cm]{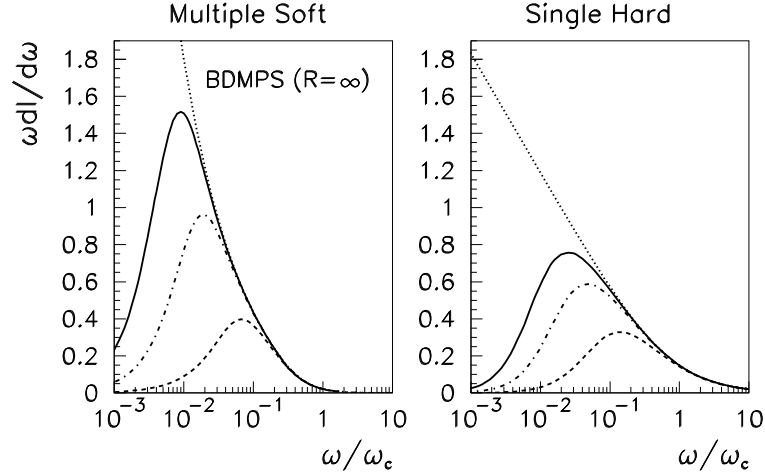}
\caption[a]{{
The medium-induced gluon energy distribution
$\omega \frac{dI}{d\omega}$ in the multiple soft scattering
approximation for different values of the
kinematic constraint $R = \omega_c\, L$=1000, 10000 and 40000.
}}
\label{figdidw}
\end{center}
\end{figure}

For practical applications, the $k_t$-integrated spectrum is needed. The
kinematical limits $0<k_t<\omega$ are imposed to compute the spectra
of Fig. \ref{figdidw} for different values of $R=\omega_c L$. A comparison
with the BDMPS result is also shown.
The origin of
R is simple: 
as $k_t^2$ is limited by $\hat q L$,
the upper kinematical limit in the $k_t$-integration
cuts the gluon energies $\omega^2\lsim  k_t^2 \sim \hat q L$. So, 
the spectrum is suppressed for
\begin{equation}
\left(\frac{\omega}{\omega_c}\right)^{2}\lsim\frac{2}{R}\ .
\end{equation}
The position of the maxima is in agreement with this estimate. Thus,
the suppression in the soft part of the spectra 
can be understood by formation time arguments. In this way, the
fact that the radiation spectrum shows small sensitivity to the infrared region
is ground on general arguments rather than 
on the actual realization of the model.
This has important consequences in the experimental observables as we will see.
The limit
$R\to \infty$ is obtained by integrating the spectrum in  $k_t^2<\infty$,
\begin{equation}
  \lim_{R\to \infty}\,
   \omega \frac{dI}{d\omega} = \int_0^\infty dk_t^2
\omega\frac{dI}{d\omega dk_t^2}
   = \frac{2\alpha_s C_R}{\pi}\, {\rm Re} \left[
   \,\ln \left (
   {\cos\sqrt{\frac{\omega_c}{i\omega}}}
\right )\right ]
   \label{eqbdmps}
\end{equation}
which coincides with the BDMPS result\cite{bdmps}. 
The average parton energy loss is, then,
\begin{equation}
 \langle \Delta E \rangle \equiv
  \int_0^\infty d\omega\,
   \omega \frac{dI}{d\omega}\hspace{0.2cm}
\xrightarrow[R\to\infty]{}
   \hspace{0.2cm}
  \frac{\alpha_s C_R}{2}\, \omega_c\, \propto\  \hat q L^2\, .
 \label{eqdeltaE}
\end{equation}
This is the well-known $L^2$ dependence of the average radiative energy loss 
\cite{bdmps,z,glv,Wiedemann:2000za,Salgado:2003gb}.

\subsection{Expanding medium}

The medium produced in a heavy ion collision is expanding very fast in the
longitudinal and probably also in the transverse direction. 
The expansion is usually parametrized
by an exponential decrease of the medium density as $n(\tau)\sim
1/\tau^\alpha$, with $\alpha$=1 for 1-dimensional (Bjorken) expansion.   
In this case, the transport coefficient changes accordingly as 
$\hat q(\tau)=\hat q_0 (\tau_0/\tau)^\alpha$ and the corresponding spectrum
can be obtained from eq. (\ref{eqspec}). It has been found in Ref. 
\cite{Salgado:2002cd}
that any expanding medium can be related with a equivalent static one with
effective transport coefficient 
\begin{equation}
  \bar{\hat{q}} = \frac{2}{L^2} \int_{\xi_0}^{L+\xi_0} d\xi\,
  \left(\xi - \xi_0\right)\, \hat{q}(\xi)\, .
\label{eqscal}
\end{equation}
This result confirms previous relations to the level of the average $\Delta E$
\cite{Baier:1998yf,Gyulassy:2000gk}
and allows to use the static formulas for any expanding scenario.

\section{Applications}

\begin{figure}[h]
\begin{center}
\includegraphics[width=10cm]{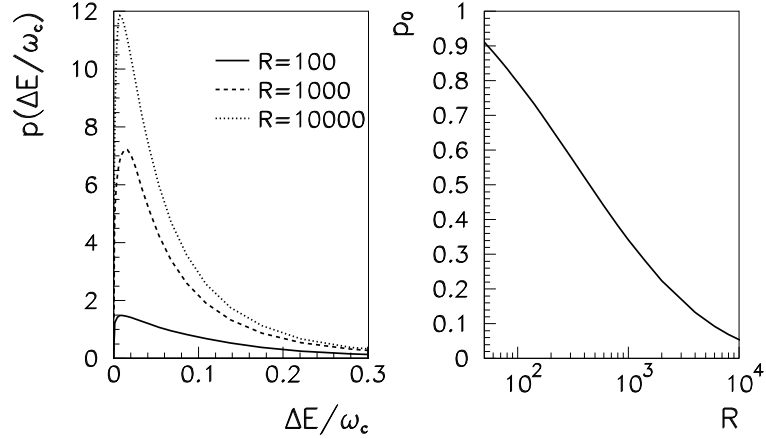}
\end{center}
\caption{
The two contributions to
the probability (\ref{eqqw}) that a parton loses $\Delta E$
of its energy in the medium: Continuous part
(left panel) and the discrete probability $p_0$ in (\ref{eqqw2})
that the hard parton
escapes the medium without interaction (right panel).
}
\label{figqw}
\end{figure}

Equation (\ref{eqspec}) has been calculated for the one-gluon inclusive case.
Up to now no progress has been made in computing diagrams with more
than one gluon emission, so, for practical applications one usually assumes
the independent gluon emission approximation \cite{Baier:2001yt}
%
\begin{equation}
  P_E(\epsilon) = \sum_{n=0}^\infty \frac{1}{n!}
  \left[ \prod_{i=1}^n \int d\omega_i \frac{dI(\omega_i)}{d\omega}
    \right]
    \delta\left(\epsilon -\sum_{i=1}^n {\omega_i \over E} \right)
    \exp\left[ - \int d\omega \frac{dI}{d\omega}\right]\ .
\label{eqqw}
\end{equation}
In the case of small gluon multiplicities, the
interference terms are expected to be
small \cite{Baier:2001yt}, and (\ref{eqqw}) should give a good approximation.
For
a medium of finite length $L$, there is a finite probability $p_0$
that no energy is lost -- no gluon is radiated and 
the fragmentation is not 
affected. This
discrete contribution decreases with increasing in medium path-length
or increasing density of the medium. So, we write 
\begin{equation}
P_E(\epsilon)=p_0\delta(\epsilon)+p(\epsilon)\, .
\label{eqqw2}
\end{equation}
In Fig. \ref{figqw} we plotted the
discrete, $p_0$, and continuous, $p(\epsilon)$, contributions 
to the {\it quenching weights}, $P_E(\epsilon)$ 
for different values of $R=\omega_c\ L$.

\subsection{Inclusive particle production}

\begin{figure}
\begin{center}
\includegraphics[width=7cm]{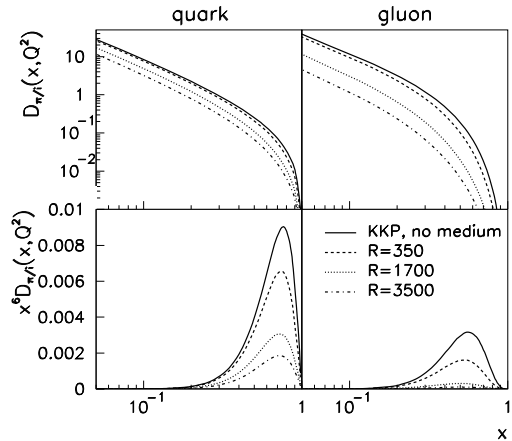}
\caption{Fragmentation functions for quarks and gluons into $\pi$ for media
of different $R=\omega_c L$. The vacuum fragmentation functions are taken 
from~\protect\cite{Kniehl:2001}. 
}
\label{figff}
\end{center}
\end{figure}

For high enough $p_t$ of the parton, the hadronization  takes place outside 
the medium. In this case, the medium-modified
fragmentation function is usually written as 
\cite{Wang:1996yh,Gyulassy:2001nm,Salgado:2002cd}
\begin{equation}
  D_{k\to h}^{(\rm med)}(z,Q^2) = \int_0^1 d\epsilon\, P_E(\epsilon)\,
  \frac{1}{1-\epsilon}\, D_{k\to h}(\frac{z}{1-\epsilon},Q^2)\, ,
  \label{eqff}
\end{equation}
\noindent
where $D_{k\to h}$ is the vacuum fragmentation function.
The only effect of the medium in (\ref{eqff}) is a shift in the energy of
the initial parton given by $P_E(\epsilon)$. Additional (logarithmic)
modifications in the $Q^2$-dependence are neglected, as they are subdominant as
compared to $\epsilon=\Delta E/E\sim 1/Q$ (we identify $Q$ with the 
initial transverse energy of the parton $E$).
In Fig. \ref{figff} the fragmentation
functions for different media computed by (\ref{eqff})
are compared to the corresponding vacuum case \cite{Salgado:2002cd}.
%

\begin{figure}
\begin{minipage}{6cm}
\begin{flushleft}
\includegraphics[width=6cm]{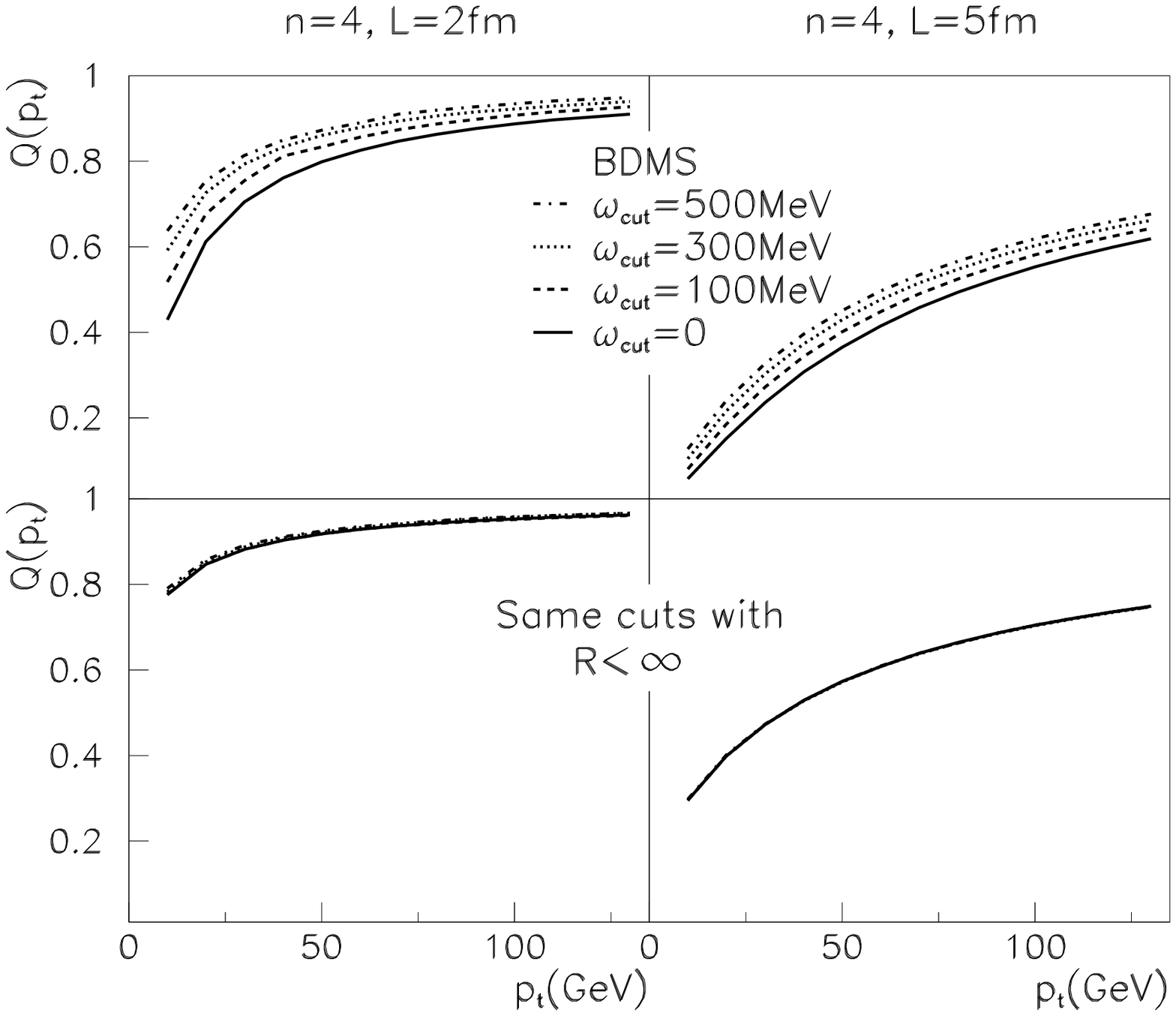}
\end{flushleft}
\end{minipage}
\vspace{\fill}
\hspace{\fill}
\begin{minipage}{6cm}
\begin{flushright}
\includegraphics[width=6cm]{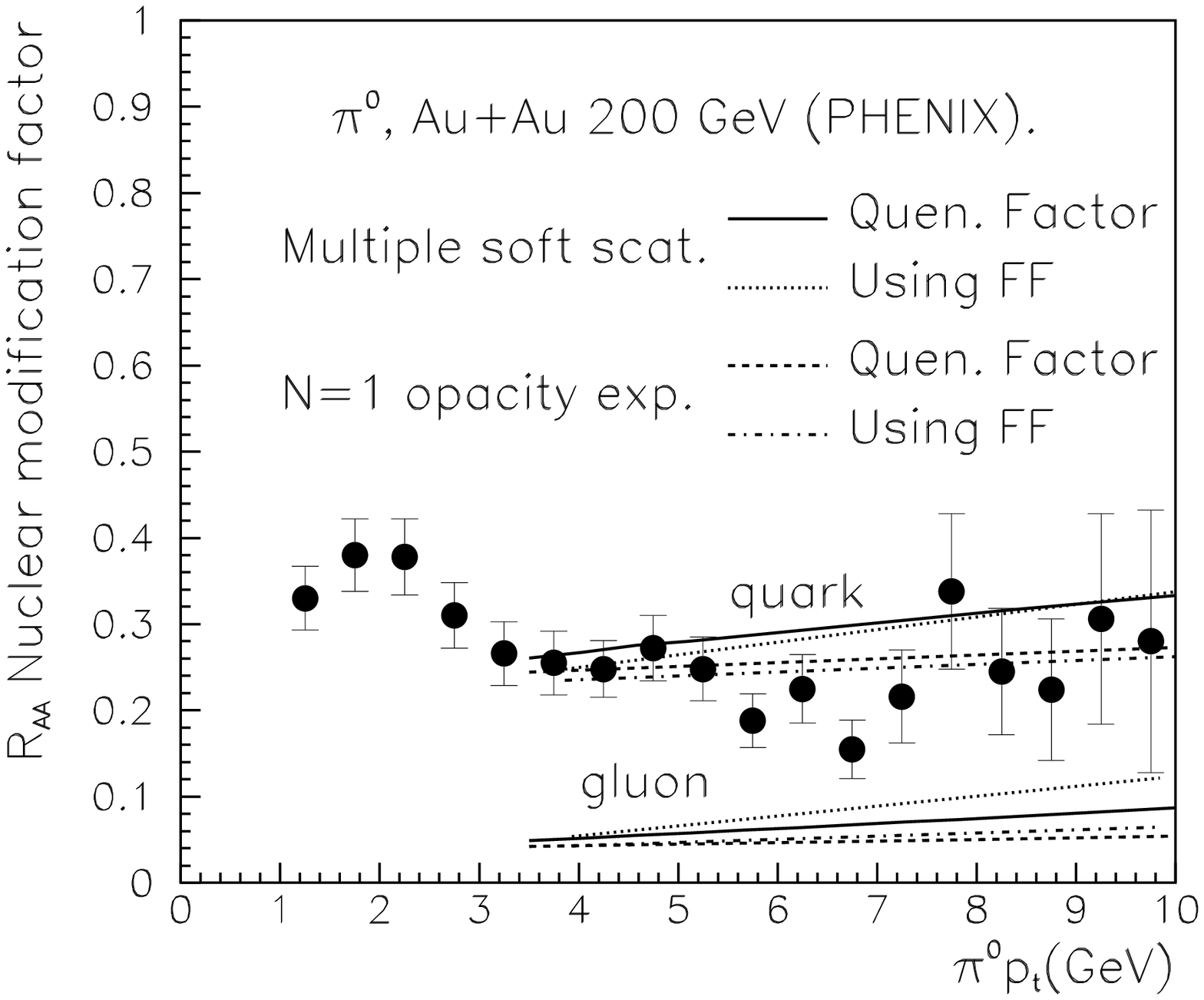}
\end{flushright}
\end{minipage}
\caption{LHS: Quenching factors (\protect\ref{eqqf})
computed from the BDMPS spectrum 
(upper figures) and with finite R (lower figures) applying
different cuts to the small-$\omega$ region ($\hat q$=1 GeV$^2$/fm 
has been taken).
 RHS: Comparison of the
suppression obtained in the multiple soft and in the
single hard scattering approximations with the experimental data from PHENIX
\protect\cite{Adler:2003qi} ($\omega_c$=67.5 GeV and R=2000 for
both multiple soft and single hard scattering approximations).
}
\label{figrhic}
\end{figure}

In order to estimate the suppression of produced $\pi$'s we make use
of the observation \cite{Eskola:2002kv} that the partonic cross section and the
PDF essentially contributes with $z^6$ to the integral in (\ref{eqpqcd}). 
In Fig. \ref{figff} we weight the fragmentation function by this factor, the
ratios at the maxima between a medium and the vacuum gives the corresponding 
suppression of final $\pi$'s. 
A suppression by a factor of $4\div 5$, as measured at RHIC, can be reached 
for R=1000 $\div$ 2000 (see also Fig. \ref{figrhic}). These values are
in agreement\cite{Salgado:2002cd} with the ones obtained by the GLV group
\cite{Gyulassy:2000gk}.

In order to study the sensitivity of these results to the small $\omega$-region,
the so-called  {\it quenching factors} have been introduced in Ref.
\cite{Baier:2001yt},
\begin{eqnarray}
 Q(p_t)&=&
{{d\sigma^{\rm med}(p_t)/ dp^2_t}\over
{d\sigma^{\rm vac}(p_t)/ dp^2_t}}=
\int d{\Delta E}\, P(\Delta E)\left(
{d\sigma^{\rm vac}(p_t+\Delta E)/ dp^2_t}\over
{d\sigma^{\rm vac}(p_t)/ dp^2_t}\right)\, ,
 \label{eqqf}
\end{eqnarray}
where the vacuum spectrum is usually taken as 
$d\sigma^{\rm vac}(p_t)/ dp^2_t\sim p_t^{-n}$.
This can be seen as an alternative way of computing the effects of jet 
quenching.
The sensitivity of the results to the infrared region can be studied 
by cutting-off the spectrum for $\omega\leq\omega_{\rm cut}$ and computing
 (\ref{eqqf}) -- see Fig. \ref{figrhic}. 
A strong sensitivity to the small-$\omega$ region appears when
the BDMPS spectrum (\ref{eqbdmps}) is used \cite{Baier:2001yt,Salgado:2003gb}.
With the regularization of the small$-\omega$ region due to finite $R=\omega_c
L$, this sensitivity practically disappears  \cite{Salgado:2003gb}.
In the RHS of Fig. \ref{figrhic} a comparison with PHENIX data 
of the suppression of $\pi^0$ for the most central AuAu collisions at RHIC
is performed. 
The magnitude and the slope
of the effect is in agreement with the data.
This is in contrast with previous expectations based on BDMPS spectrum
of a much steeper slope, see Fig. \ref{figrhic} (LHS). In this way, the results
for the multiple soft and the single hard scattering approximations are
similar.
%

\subsection{Jet shapes}

Equation (\ref{eqspec}) relates the energy loss of a parton with the transverse
momentum broadening of the associated gluon radiation. This dynamics should 
modify the internal jet substructure from the vacuum case.
In order to study these effects, a first attempt to compute jet observables
in the presence of a medium
has been made in \cite{Salgado:2003rv}. In the rest of the section we present 
the medium--modification for two quantities,
the fraction of jet energy inside a cone and
the gluon multiplicity distribution.

\begin{figure}[h]\epsfxsize=8.5cm
\centerline{\epsfbox{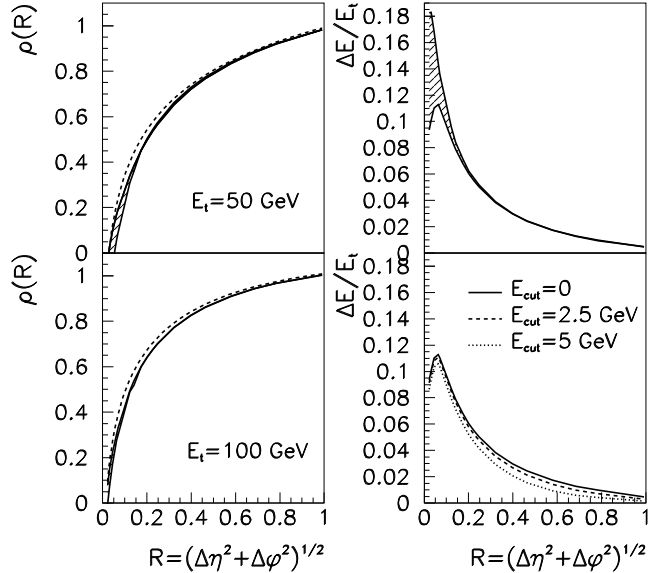}}
\caption{LHS: The jet shape (\protect\ref{eq5}) for a 50 GeV and
100 GeV quark-lead jet which fragments in the vacuum 
\protect\cite{D0} (dashed curve) or
in a dense QCD medium (solid curve)
characterized by $\omega_c = 62$ GeV and $\omega_c\, L = 2000$.
RHS: the corresponding
average medium-induced energy loss for $E_t = 100$ GeV
outside a jet cone $R$ radiated
away by gluons of energy larger than $E_{\rm cut}$. Shaded regions
indicate theoretical uncertainties discussed in the text.
}\label{figfig2}
\end{figure}

The fraction of the jet energy inside a cone of radius 
$R=\sqrt{(\Delta \eta)^2 + (\Delta \Phi)^2}$ is
\begin{equation}
  \rho_{\rm vac}(R) = \frac{1}{N_{\rm jets}} \sum_{\rm jets}
  \frac{E_t(R)}{E_t(R=1)}\, .
  \label{eq5}
\end{equation}
\noindent
In the presence of the medium, this energy is shifted by\cite{Salgado:2003rv}
\begin{equation}
  \rho_{\rm med}(R) =
    \rho_{\rm vac}(R) - \frac{\Delta E_t(R)}{E_t}
        + \frac{\Delta E}{E_t} \left( 1 - \rho_{\rm vac}(R)\right)\, ,
      \label{eq9}
\end{equation}
where $\Delta E_t(R)$ is the additional (medium) energy radiated 
outside a cone $\Theta=R$ and
$\Delta E(\Theta)=\int \epsilon P(\epsilon,\Theta)d\epsilon$,
where the quenching weight is computed by
integrating the spectrum (\ref{eqspec}) in
$\omega\sin\Theta < k_t < \omega$.
In Fig. \ref{figfig2} we plot the medium-shifted distributions. The shaded area 
corresponds to the uncertainty in finite quark-energy effects: in the
eikonal approximation
$P(\epsilon)$ have support in the unphysical region $\epsilon > 1$.
To estimate this effect we make the change
$P(\epsilon)\,\rightarrow\, P(\epsilon)/\int_0^1 d\epsilon P(\epsilon)$.
%
%
%

The effect of the medium is very small (at $R$=0.3, it is $\sim$ 5\% for 
a 50 GeV jet and $\sim$ 3\% for a 100 GeV jet). The smalleness of this effect
could allow for a calibration of the total energy of the jet
without tagging in a recoiling hard photon or Z-boson. It also implies
that the jet $E_t$ 
cross section scales with the number of binary collisions.
In order to check the sensitivity of our results to the small-energy region,
we impose, in analogy to the previous section,  low momentum cut-offs
which removes gluon emission below 5 GeV. It is interesting that
transverse momentum 
broadening is very weakly affected by these cuts. This is again
due to 
the infrared behavior of the spectrum for small values of $\omega$ 
-- see Fig. 
\ref{figdidw}. A proper substraction of the large background present
in heavy ion collisions would benefit from this result.

\begin{figure}[t!]\epsfxsize=8cm
\centerline{\epsfbox{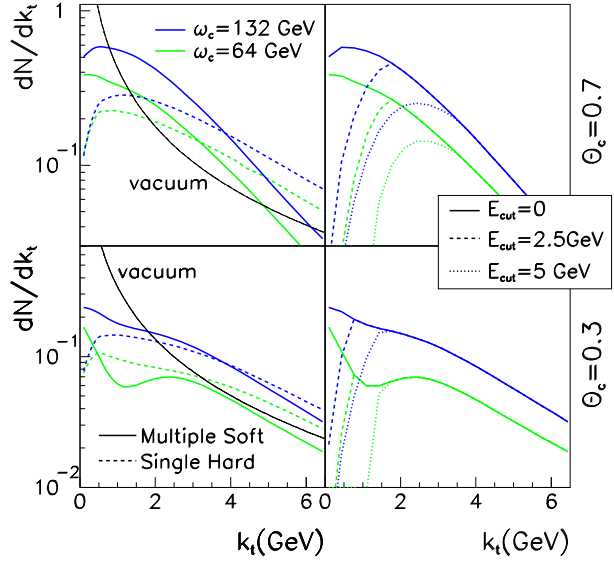}}
\caption{Comparison of the vacuum and medium-induced part of the
gluon multiplicity distribution
(\protect\ref{eq8}) inside a cone size $R=\Theta_c$, measured as a
function of $k_t$ with respect to the jet axis. Removing gluons with
energy smaller than $E_{\rm cut}$ from the distribution (dashed and dotted
lines) does not affect the high-$k_t$ tails.
}\label{figglmult}
\end{figure}

$k_t$-differential measurements are expected to be more sensitive to 
medium effects. As an example, 
the intrajet multiplicity of produced gluons as
a function of the transverse (with respect to the jet axis) momentum
is plotted in Fig. \ref{figglmult}.
The medium-induced additional number of gluons
with transverse momentum $k_t = \vert {\bf k}\vert$, produced
within a subcone of opening angle $\theta_c$, is
\begin{eqnarray}
 \frac{dN_{\rm med}}{dk_t} =  \int_{k_t/\sin\theta_c}^{E_t} d\omega\,
               \frac{dI_{\rm med}}{d\omega\, dk_t}\, .
  \label{eq8}
\end{eqnarray}
For the vacuum we simply assume
$dN_{\rm vac}/dk_t\sim 1/k_t\log(E_t\sin\theta_c/k_t)$.
In this case, 
the effect is sizable for transverse momenta of the order of several GeV
and could be easily measured experimentally.
A more realistic analysis 
would need of an implementation of the whole fragmentation.
However, the origin of the shift is mainly due to the large $k_t\sim 
Q_{\rm sat}$ that the gluon obtains from the medium. In this
way, we expect this conclusion to be very robust and not depending on the
actual realization of the model.

\section{Discussion and other approaches}
\label{secoutros}

In the previous sections, we have presented the usual framework to
compute high-$p_t$ particle production in nuclear collisions. It is based on 
the collinear factorization, eq. (\ref{eqpqcd}), 
supplemented with nuclear parton distribution
functions and final state effects due to medium-induced gluon radiation. 
This framework has been successfully employed to reproduce experimental data. 
Let us comment on the differences with other approaches within the same 
framework, concerning both the initial and the
final state. 

First, in the initial state 
intrinsic-$k_t$ and/or Cronin effect due to multiple (elastic) scattering
are sometimes introduced (see e.g. 
\cite{Vitev:2002aa,Levai:2003at,Vitev:2002pf}). The
$p\bar p$ data cannot distinguish between these two approaches, however, $hA$ 
data at energies of several tens of GeV are well described with this
mechanism. Its magnitude for central AuAu collisions at RHIC can be as large
as a 50\% increase. This increase is sometimes compensated by the large 
shadowing of Ref. \cite{Li:2001xa}, however, as we have seen this strong
gluon shadowing is in disagreement with DIS data. The nuclear PDF obtained
in a DGLAP analysis result only in a tiny enhancement 
\cite{Eskola:2002kv,deFlorian:2003qf}.

Concerning the final state effects, most of the approaches rely on the 
radiative energy loss and differ only on the approximation used, multiple
soft \cite{bdmps} or single hard scattering \cite{glv}. We have seen that both
approximations give very similar results when the appropriate kinematical limits
and correspondingly similar parameters are taken into account
\cite{Salgado:2003gb}. (For an approach based on twist expansion
in DIS see Ref. \cite{Guo:2000nz}). The possibility of collisional energy loss
has been also explored \cite{Mustafa:2003vh} with a reasonable result. As
it has been exposed in Section 3, formation time arguments lead to a
radiative energy loss which increases as $L^2$, in the case of a collisional
energy loss the growth is, however, as $L$. So, the centrality dependence of the
effect is expected to be sensitive to this different behavior. Unfortunately,
it seems that present
data cannot distinguish between a $L$ or $L^2$ behavior
\cite{Choi:2003pq}. Notice that in all these analyzes, hadronization is 
assumed to take place outside the medium. This could not be the case for
the smallest $p_t$ values \cite{Arleo:2002kh}.

On the other hand, 
an initial state origin for the suppression of high-$p_t$ particle 
production has been proposed in the framework of the saturation approach 
\cite{Kharzeev:2002pc}, the origin being the smaller number of initial gluons
in the nuclei wave functions. There has been some discussion on
whether this removes the Cronin enhancement at intermediate values
of $p_t$ or not, but finally the different groups 
agree \cite{Albacete:2003iq,Baier:2003hr,Kharzeev:2003wz,Jalilian-Marian:2003mf}
in that saturation leads to a suppression for all values of $p_t$.
$dAu$ experimental data ruled out this hypothesis as the main 
source of high-$p_t$ particle suppression at central rapidities. 
However the prediction is 
\cite{Albacete:2003iq} that at higher energies and/or rapidities this mechanism
very efficiently suppresses the high-$p_t$
particle yields. The new preliminary data
from BRAHMS \cite{brahmsprel} 
find a strong reduction of $\pi$'s for $p_t < 2.5$ GeV in the forward
direction. From the results in Section 2, a suppression like this 
seems difficult
to accommodate in a DGLAP approach in collinear factorization.
If this preliminary data is confirmed it could be the first clear indication
of saturation phenomena in nuclear collisions.

\section{Conclusions}

In this review we have described the most recent theoretical results
relating medium-induced gluon radiation, energy loss and jet broadening. 
These effects are accessible for the first time in experiments of heavy ion
collisions at RHIC. All the experimental data strongly point to 
a large jet suppression due to interaction
with the produced medium. The larger
energy of the LHC will allow for a qualitative new regime, where the 
jets are not completely suppressed and the jet substructure could be 
measured in the large background environment. 
This will open a completely new window for the study of the evolution of 
high energetic particles in a medium.

{\bf Acknowledgments:} 
I would like to thank J. Albacete, N. Armesto, K. Eskola, H. Honkanen,
V. Kolhinen, A. Kovner, V. Ruuskanen and U. Wiedemann for the very nice
and fruitful collaboration which is partially reviewed in this paper. 
Critical reading of this manuscript 
by N. Armesto and U. Wiedemann is gratefully acknowledged.

\end{document}